\documentclass[a4paper,conference]{IEEEtran}
\ifCLASSINFOpdf
\else
\fi
\usepackage{graphicx}
\usepackage{float}
\usepackage{setspace}
\usepackage{amsmath}
\usepackage{booktabs}
\usepackage{amssymb}
\usepackage{pifont}
\newcommand{\setParDis}{\setlength {\parskip} {0.3cm} }
\newcommand{\setParDef}{\setlength {\parskip} {0pt} }
\begin{document}
%
\title{Multi-modal Brain Tumor Segmentation via Missing Modality Synthesis and Modality-Level Attention Fusion}

\makeatletter
\newcommand{\linebreakand}{%
  \end{@IEEEauthorhalign}
  \hfill\mbox{}\par
  \mbox{}\hfill\begin{@IEEEauthorhalign}
}
\makeatother

\author{
  \IEEEauthorblockN{Ziqi Huang}
\IEEEauthorblockA{Department of Electronic and\\Electrical Engineering\\
Southern University of Science\\and Technology\\
Email: huangzq@mail.sustech.edu.cn}
  \and
  \IEEEauthorblockN{Li Lin}
\IEEEauthorblockA{Department of Electronic and\\ Electrical Engineering\\
Southern University of Science\\and Technology\\
The University of Hong Kong\\
Email: linli@eee.hku.hk}
  \and
  \IEEEauthorblockN{Pujin Cheng}
\IEEEauthorblockA{Department of Electronic and\\Electrical Engineering\\
Southern University of Science\\and Technology\\
Email: 12032946@mail.sustech.edu.cn}
  \linebreakand 
    \IEEEauthorblockN{Linkai Peng}
    \IEEEauthorblockA{Department of Electronic and\\Electrical Engineering\\
    Southern University of Science\\and Technology\\
    Email: lkpengcs@gmail.com}
  \and
  \IEEEauthorblockN{Xiaoying Tang*}
    \IEEEauthorblockA{Department of Electronic and\\Electrical Engineering\\
    Southern University of Science\\and Technology\\
    Corresponding Author Email: tangxy@sustech.edu.cn}
}


%


\maketitle

\begin{abstract}
Multi-modal magnetic resonance (MR) imaging provides great potential for diagnosing and analyzing brain gliomas. In clinical scenarios, common MR sequences such as T1, T2 and FLAIR can be obtained simultaneously in a single scanning process. However, acquiring contrast enhanced modalities such as T1ce requires additional time, cost, and injection of contrast agent. As such, it is clinically meaningful to develop a method to synthesize unavailable modalities which can also be used as additional inputs to downstream tasks (e.g., brain tumor segmentation) for performance enhancing. In this work, we propose an end-to-end framework named Modality-Level Attention Fusion Network (MAF-Net), wherein we innovatively conduct patchwise contrastive learning for extracting multi-modal latent features and dynamically assigning attention weights to fuse different modalities. Through extensive experiments on BraTS2020, our proposed MAF-Net is found to yield superior T1ce synthesis performance (SSIM of 0.8879 and PSNR of 22.78) and accurate brain tumor segmentation (mean Dice scores of 67.9\%, 41.8\% and 88.0\% on segmenting the tumor core, enhancing tumor and whole tumor). 
\end{abstract}


%
\IEEEpeerreviewmaketitle

\section{Introduction}
Glioma is the most common intracranial tumor. Currently, magnetic resonance (MR) imaging is the best observation approach for diagnosing and evaluating glioma before surgery. In MR imaging, due to different settings of scanning parameters such as the radio frequency pulse, gradient field and signal acquisition time, MR modalities of different contrasts can be generated. Typically, a combination of different MR modalities benefits more accurate disease diagnosis. In clinical practice, general MR modalities such as T1-weighted, T2-weighted, and fluid attenuation inversion recovery (FLAIR) can be relatively easily obtained. These modalities do not require external interference and can be acquired through adjusting the scanning parameters. However, there are also modalities that are relatively difficult to obtain, such as contrast-enhanced T1 (T1ce); specific contrast agent, which may cause allergic or adverse reactions, is injected into a patient's body to highlight certain brain tissues. Moreover, collecting T1ce requires extra time and cost. 


Existing evidence \cite{miccai2019dou,tip2021missing,hamghalam2021modality} suggests that T1ce plays an important role in identifying the tumor core and the enhancing tumor of glioma, as it can highlight the active tumor and clearly distinguish the boundary between necrotic and parenchyma areas. According to neurosurgeons, the intensity and boundary of active tumors are important biomarkers for subsequent tumor grading and treatment-related decision-making. In such context, T1ce offers indispensable information for glioma diagnosis even though it is highly likely to be a missing modality in real clinics. On the other hand, previous works have suggested \cite{cvpr2021translate} that image segmentation tasks and image generation tasks are typically mutually beneficial, although this hypothesis has not been verified on the tumor segmentation task. In light of this, we conjecture that synthesizing the T1ce modality will also benefit the subsequent brain tumor segmentation task.

In this work, we introduce an innovative framework called Modality-Level Attention Fusion Network (MAF-Net) for brain tumor segmentation. Our main contributions are three-fold: 
\begin{itemize}
    \item We propose the first multi-modal patchwise contrast learning framework for modality synthesis and apply it to synthesizing T1ce.
    \item Multi-modal MR images are fully explored and integrated through a modality-level attention fusion block.
    \item The modality synthesis and tumor segmentation tasks are performed simultaneously to promote each other's performance, and their feasibility and effectiveness have been established through extensive experiments. 
\end{itemize}

\begin{figure*}[htbp]
\centering
\includegraphics[width=16cm]{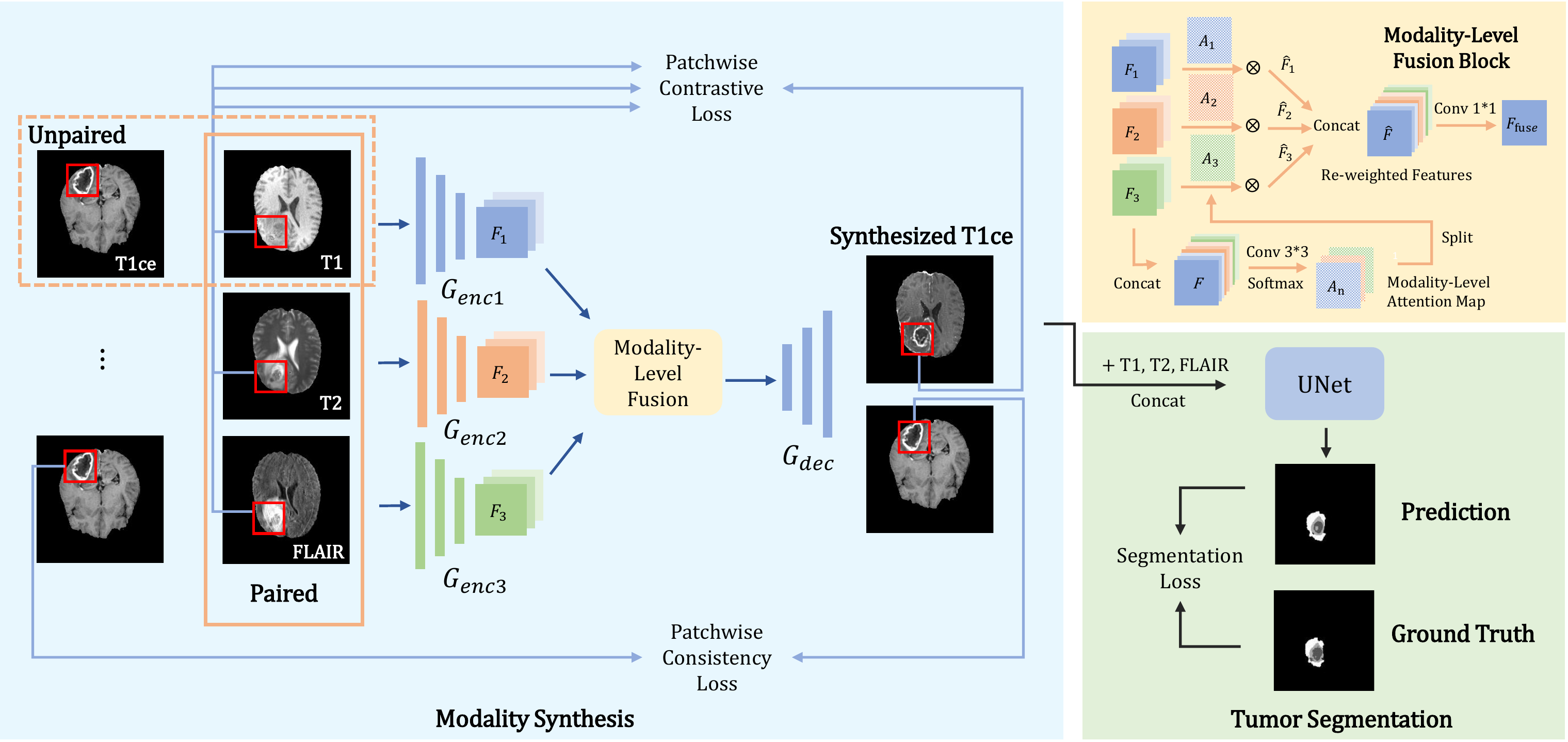}
\vspace{-0.25cm}
\caption{Flowchart of our proposed MAF-Net for missing modality synthesis and tumor segmentation. The discriminator $D$ is not shown in the schematic diagram.} 
\label{fig1} 
\end{figure*}

\section{Related Works}

\textbf{Brain Tumor Segmentation. }
Numerous brain tumor segmentation studies employ multi-modal MR images to pursue superior results. Traditional methods \cite{multimodal1} use generative probabilistic models to deal with multi-modal information. With the rapid development of deep learning, neural network related tumor segmentation models \cite{deepmultimodal1,deepmultimodal2} have emerged, yielding great success. For example, Hua et al. \cite{deepmultimodal4} uses cascaded networks to utilize the multi-modal information, whereas Ranjbarzadeh et al. \cite{deepmultimodal3} combines cascaded networks with an attention mechanism and produces more accurate predictions.

\textbf{Image-to-image translation. }
In computer vision, image-to-image translation has been a very popular research topic. Most image generation and translation methods build their basis on generative adversarial network (GAN) \cite{goodfellow2014generative}. For example, Zhu et al. proposes Pix2pix \cite{pix2pix} to effectively solve the paired image translation task, and CycleGAN \cite{Zhu2017UnpairedIT} applies a cycle consistency loss to the task of unpaired image translation with a strong bijective constraint. Further, CUTGAN \cite{cutgan} introduces patchwise contrastive learning, mining high-dimensional features by constructing positive and negative samples at a patch level. 

\textbf{Multi-modal Image Fusion. }
To deal with multi-modal images, a fusion step that can effectively combine the information of each modality is essential for downstream tasks. Zhou et al. \cite{fusion2} carries out fusion at a decision level, training multiple SVM \cite{SVM} classifiers for prediction ensemble. 
On the other hand, Chen et al. designs a dual-task network \cite{fusion1} conducting efficient fusion at a high-dimension feature level. Meanwhile, SE-Net \cite{SEmodule} introduces a feature level fusion module named SE module to enhance essential features and reduce the importance of task irrelevant features, obtaining attention-guided features for fusion. Recently, self-attention is explored to accomplish multi-modal fusion \cite{transfuse} at the cost of extensive computation. 

\textbf{Multi-modal Medical Image Synthesis. }
Several GAN-based image generation methods have also been applied to multi-modal medical image settings. For example, \cite{drit_use_in_medical} and \cite{TMI2020} disentangle images into content maps and style codes and recombine them to form style-transferred images. However, they do not work for brain MRI given that the intensity distribution within a single MRI modality is too diverse. Previous researches  \cite{miccai2019dou,tip2021missing,isbi2020} use modality generation or reconstruction as an auxiliary task to complement the missing MR modalities, and adopt the generated images or latent features for downstream tasks. Nevertheless, these methods do not perform well on synthesizing T1ce nor segmenting brain tumor, and they typically require paired T1ce samples at least during training. Due to the scarcity of T1ce, sometimes we have access to only a small amount of T1ce but a large number of common modality MR images, resulting in mismatch between T1ce and general MR modalities. To accommodate such situation, unpaired training frameworks are needed.

\section{Method}

\label{sec:format}
Given a set of paired regular MR modalities including T1, T2 and FLAIR and a set of unpaired T1ce, it is desirable that a model can generate reasonable T1ce images which accommodate well a tumor segmentation network together with real T1, T2, and FLAIR images. Our proposed MAF-Net pipeline mainly consists of two modules: a missing modality synthesis module and a tumor segmentation module. Fig. \ref{fig1} demonstrates the overall architecture of MAF-Net.

\vspace{0.05cm}
\textbf{Unpaired Missing Modality Synthesis via Patchwise Contrastive Learning. }
The unpaired missing modality synthesis task aims to utilize unpaired dataset, learn correspondences and differences between source and target modalities, and eventually complement the missing modality’s information. We use CUTGAN as our baseline for the missing modality synthesis task. 

Keeping consistent with CUTGAN, we follow the patch-wise contrastive framework, but expand it to three encoders and one decoder; each encoder focuses on learning the relationship of intensity and structure between a specific modality of interest (T1, T2, or FLAIR) and the to-be-generated target modality T1ce. After identifying the common features between each modality and T1ce, those features are fused and fed into the decoder to generate the synthesized T1ce. A GAN-based framework includes a discriminator which tries to distinguish whether the current image is a real T1ce or a synthesized one, and a generator which learns synthesizing real-like T1ce images. The objective function $L_{syn}$ for the modality synthesis task is
\begin{spacing}{1}
\vspace{-0.5cm}
\begin{small}
\begin{equation}
    \begin{aligned}
    \label{eq1}
        L_{syn} = L_{GAN}(G,D,X,Y)&+\frac{1}{N} \sum_{i=1}^{N}\lambda _{X}L_{PatchNCE}^{X_i}(G,H,X) \\
                            &+\lambda _{Y}L_{PatchNCE}^Y(G,H,Y),
    \end{aligned}
\end{equation}
\end{small}where $L_{GAN}$ represents the adversarial loss

\begin{small}
\begin{equation}
    \begin{aligned}
    \mathcal{L}_{\mathrm{GAN}}(G, D, X, Y) &=\mathbb{E}_{\boldsymbol{y} \sim Y} \log D(\boldsymbol{y}) \\
    &+\mathbb{E}_{\boldsymbol{x} \sim X} \log (1-D(G(\boldsymbol{x}))).
    \end{aligned}
\end{equation}
\end{small}It encourages the distribution of the fake image to be close to that of the real one. $N$ denotes the total number of the regular modalities and is three in our task. $X=\left \{  X_{i} \right |i=1,2,3\}$ denotes a set of paired MR images of three different modalities, and $Y$ is a set of unpaired T1ce. $G$ and $D$ respectively represent the generator and the discriminator. We set $\lambda_X = 1$ and $\lambda_Y = 1$ when we jointly train the consistency loss $L_{PatchNCE}^Y$; otherwise we have $\lambda_X = 10$ and $\lambda_Y = 0$. $L_{PatchNCE}^{X_i}$ makes use of an MLP network $H$ to extract high-dimensional features of each input modality $X_i$ and the synthesized T1ce at a patch level, so as to calculate the NCE loss \cite{oord2018representation}. 
The features of the input images and the synthesized image are mapped into the same latent space, and the NCE loss enforces features of the synthesized T1ce stay close to those of the multiple general modalities at the co-located patch, as illustrated in Fig. \ref{patchnce}. Since we have three input modalities, the NCE loss is calculated on each modality and then gets summed up.
$L_{PatchNCE}^{X_i}$ is defined as
\setParDis
\end{spacing}
\vspace{-0.25cm}
\begin{small}
\vspace{-0.15cm}
\begin{equation}
\begin{aligned}
\label{eq2}
L_{PatchNCE}^{X_i} = E_{x\sim X_i} \sum_{l_n=1}^{L_n} \sum_{s=1}^{S_{l_n}} L_{nce}(\hat{z_{l_n}^s},z_{l_n}^s,z_{l_n}^{S/s}\mid n=i),    
\end{aligned}
\vspace{-0.15cm}
\end{equation}
\end{small}where $\hat{z_{l_n}^s}=\left \{H(G_{enc_n}^{l_n}(G(X)))  \right \}$ is the latent feature vector of the synthesized T1ce generated by layer $l$ in the $n$-th encoder at position $s \in S$; $z_{l_n}^s=\left \{H(G_{enc_n}^{l_n}(X_n))  \right \}$ represents features of the input patch and $z_{l_n}^{S/s}$ refers to features of the corresponding negative samples which are sampled from different positions in the same input image. More specifically, $L_{nce}$ is computed as
\setParDef
\begin{small}
\vspace{-0.05cm}
\begin{equation}
\begin{aligned}
\label{eq3}
L_{nce}(\hat{z_{l_n}^s},z_{l_n}^s,z_{l_n}^{S/s}) = -log\left [ \frac{e^{\hat{z_{l_n}^s}\cdot z_{l_n}^s / \tau }}{e^{\hat{z_{l_n}^s}\cdot z_{l_n}^s/\tau}+\sum_{s_i \in S/s}e^{\hat{z_{l_n}^s} \cdot z_{l_n}^{s_i}/\tau} }  \right ],
\end{aligned}
\vspace{0.1cm}
\end{equation}
\end{small}where $\tau$ is the scaling factor that is generally set to be 0.07, and $s_i$ refers to the negative patches. $L_{PatchNCE}^Y$ denotes the patchwise consistency loss in Fig. 1, comparing the synthesized T1ce image and the real one to ensure that there are no out-of-bounds abnormalities in the generated fake image.

\begin{figure}[tbp]
\centering
\includegraphics[width=9.5cm]{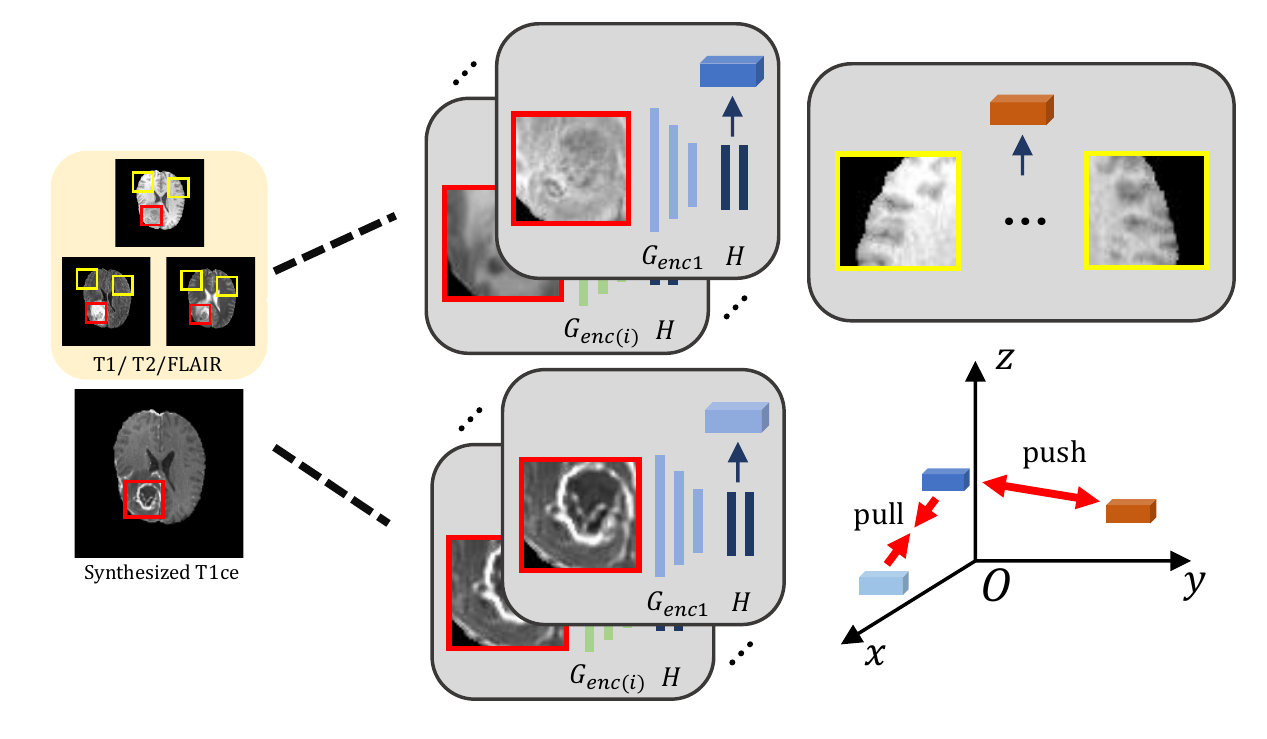}
\vspace{-0.55cm}
\caption{Demonstration of multi-modal patchwise contrastive learning. The input can be any of the general modality $X_i$ (including T1, T2 and FLAIR in our case), and $G_{enc(i)}$ refers to the corresponding encoder. The red blocks represent positive samples and the yellow blocks are negative samples. The feature extraction process for negative samples is identical with that for positive samples, which is not included in this figure.} 
\vspace{-0.5cm}
\label{patchnce} 
\end{figure}

\textbf{Dynamic Modality-Level Attention Fusion. }
Since we utilize three encoders to separately deal with the relationship between each existing modality and the target modality, we aim at deeply fusing features extracted by those different encoders. Inspired by the SE module \cite{SEmodule}, we propose Modality-Level Attention Fusion to perform dynamic modality-weighted refining and multi-modal feature fusion.
The structure of our Modality-Level Attention Fusion block (MAF-Block) is also shown in Fig. \ref{fig1}.

In the upper-right panel of Fig. \ref{fig1}, $\left \{F_n|n=1,2,3 \right\}$ are the feature vectors respectively outputted by the three encoders. We sequentially concatenate all $F_n$ together to form as $F$, feed it into a $3\times3$ convolution layer and perform activation in the modality channel, obtaining $N$ feature attention maps $A_n$ at the modality level. Each time we train the network, we assign different learnable weight maps to each input modality's features. Hypothesizing that features of all the three modalities will contribute to the synthesis of T1ce, we choose softmax as the activation function instead of sigmoid, setting the sum of the weights of the three modalities to be 1 at a patch level, to achieve the purpose of dynamic weight adjustment.
After an element-wise multiplication of $A_n$ and $F_n$, we obtain re-weighted modality-level features $\hat{F_n}$. Subsequently, we use a $1\times1$ convolution and LeakyReLU to conduct feature fusion and dimension reduction on the concatenated features $\hat{F}$, and finally get the fused features $F_{fuse}$ acting as the input of the decoder.

\textbf{Joint Learning of Synthesizing and Segmentation. }
Afterwards, we treat the generated T1ce as a supplementary modality and feed it into our brain tumor segmentation network along with the real T1, T2, and FLAIR modalities via concatenation. UNet \cite{unet} with an input channel of four is used as our tumor segmentation network and cross-entropy is chosen as the loss function. 
\begin{equation}
\begin{aligned}
\label{eq4}
L_{seg} = - \frac{1}{M} {\textstyle \sum_{i=1}^{C}}{\textstyle \sum_{j=1}^{M}} \hat{q_j^i} log(q_j^i),
\end{aligned}
\end{equation}where $C=4$ refers to the total number of tumor segmentation classes. $\hat{q_j^i}$ and $q_j^i$ respectively denote the prediction and ground truth labels at the $j$th pixel, and $M$ is the total number of pixels.

Given that the contrast enhanced tissue in T1ce is mainly the enhancing tumor which is also one of the segmentation targets, we intuitively conduct joint learning for modality synthesis and tumor segmentation. We set a weighting parameter $\lambda$ for the loss of the modality synthesis task. The overall objective function is thus
\begin{equation}
\begin{aligned}
\label{eq5}
L_{total} = \lambda L_{syn}+L_{seg},
\end{aligned}
\end{equation}where $\lambda$ is empirically set to be 1e-3, considering that brain tumor segmentation is our main task.  

\section{Results and Discussion}

\label{sec:pagestyle}

\textbf{Dataset. } The Brats2020 dataset \cite{brats} contains paired MR images of T1, T2, T1ce and FLAIR from a total of 369 glioma patients. We take T1, T2, FLAIR as the available MR modalities, and T1ce as the missing target modality for model training. Given that the axial view can display gliomas more comprehensively, we conduct T1ce synthesis and tumor segmentation in the axial view. We center crop all axial images to $224\times224$ and select slices containing the brain tumor for training. The entire dataset are divided into training, validation and test sets according to a ratio of 7:1:2 at the sample level. Each sample is manually annotated into 4 categories at the pixel level, including background, necrosis, enhancing tumor (ET) and edema areas. 
According to clinical practice, we define the union of edema, enhancing tumor, and necrosis as whole tumor (WT), and the union of necrosis and enhancing tumor as tumor core (TC). We evaluate the segmentation performance on three components, namely WT, ET, and TC, as is illustrated in Fig. \ref{tumor2}. To reduce discontinuous regions, we conduct the largest connected region operation for post-processing upon WT.

\begin{figure}[htb]
\centering
\centerline{\includegraphics[width=9.5cm]{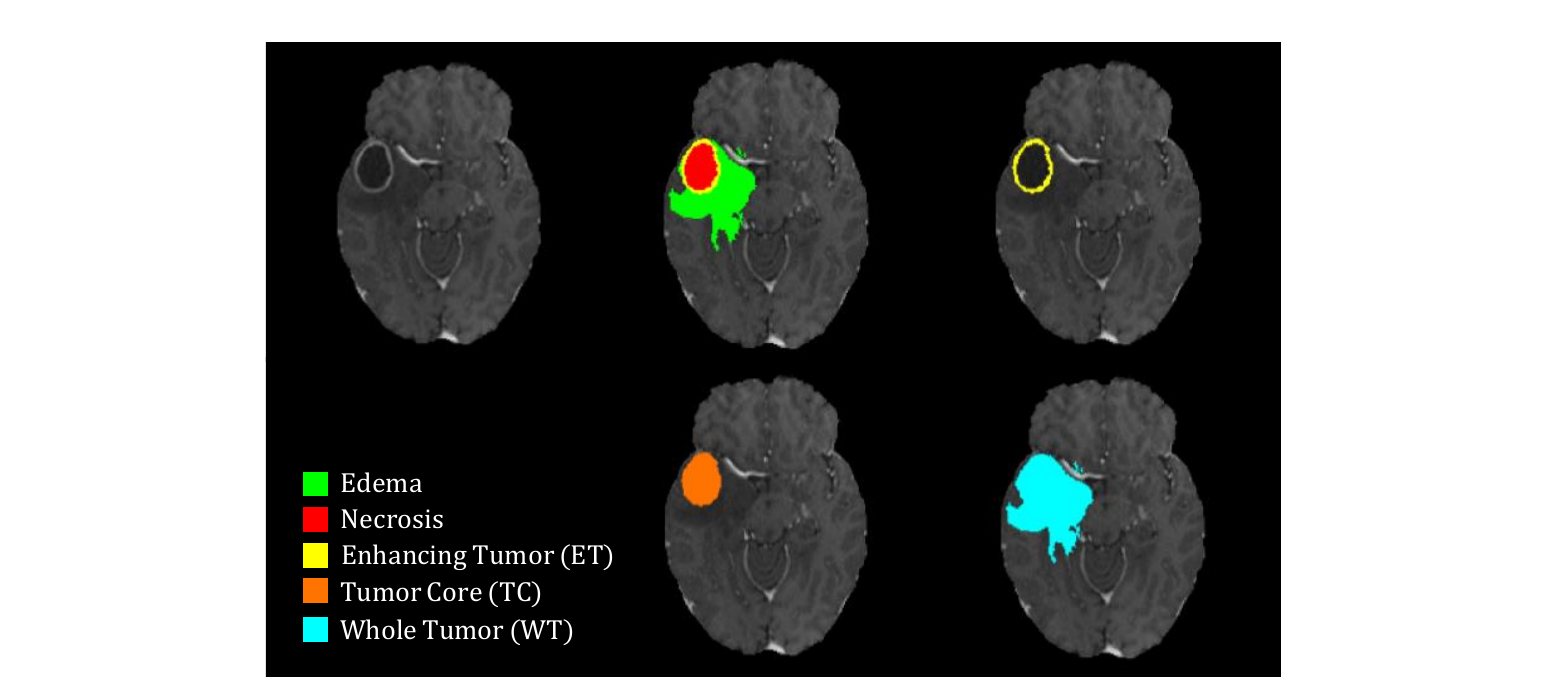}}
\vspace{-0.25cm}
\caption{Illustration of different tumor components.} 
\vspace{-0.25cm}
\label{tumor2} 
\end{figure}
\vspace{-0.0cm}

\vspace{+0.1cm}
\textbf{Training Details. }
We use a self-defined ResNet with 9 ResNet blocks and twice downsampling for the encoders and decoder of our modality synthesis module. Adam is used as the optimizer, and the batch size is set to be 4. The learning rates for the generator $G$, MLP $H$, discriminator $D$ and segmentor are respectively set to be $l_{G} = 4e-4$, $l_H= 4e-4$, $l_D = 2e-4$, and $l_{seg} = 1e-4$. In our case, the layers $L={0,4,8,12,16}$ are selected for computing the NCE loss. We first train the modality synthesis task for a total of 10 epochs, leading to stable T1ce generation. Then, we add the tumor segmentation task into the training schedule for another 100 epochs until convergence on the validation set.

\begin{table*}[t]
\centering
\caption{Ablation Analysis of Each Component in MAF-net. SingleEN denotes CUTGAN's encoder and is fed with the concatenated T1, T2 and FLAIR. MAF-GAN refers to the GAN used in the modality synthesis module and MAF-Block refers to the Modality-Level attention fusion block. Keys: WT – whole tumor, ET – enhancing tumor, TC – tumor core.}
\resizebox{\textwidth}{15.5mm}{
\begin{tabular}{ccccc|cc|cc|cc} 
\toprule
\multicolumn{5}{c|}{Component}                          & \multicolumn{2}{c|}{\textbf{WT}} & \multicolumn{2}{c|}{\textbf{ET}} & \multicolumn{2}{c}{\textbf{TC}}  \\ 
\hline
UNet & SingleEN & MAF-GAN & MAF-Block & Joint Learning & Dice$\uparrow$   & ASSD$\downarrow$           & Dice$\uparrow$   & ASSD$\downarrow$           & Dice$\uparrow$   & ASSD$\downarrow$           \\ 
\hline
\ding{51}    &          &         &            &                & 84.1\% & 5.344          & 33.9\% & 5.362          & 60.6\% & 5.858          \\
\ding{51}    & \ding{51}        &         &            &                & 85.6\% & 5.332          & 37.8\% & 4.422          & 62.5\% & 4.562          \\
\ding{51}    &          & \ding{51}       &            &                & 87.5\% & 5.990          & 35.0\% & 4.836          & 63.9\% & 4.422          \\
\ding{51}    &          & \ding{51}       & \ding{51}          &                & 87.9\% & \textbf{4.120}          & 39.9\% & 4.276          & 67.2\% & 4.453          \\
\ding{51}    &          & \ding{51}       & \ding{51}          & \ding{51}              & \textbf{88.0\%} & 4.700          & \textbf{41.8\%} & \textbf{3.845}          & \textbf{67.9\%} & \textbf{4.416}          \\
\bottomrule
\end{tabular}}
\label{ablation1}
\end{table*}

\vspace{-0.cm}
\begin{figure*}[t]
\centering
\centerline{\includegraphics[width=17.5cm]{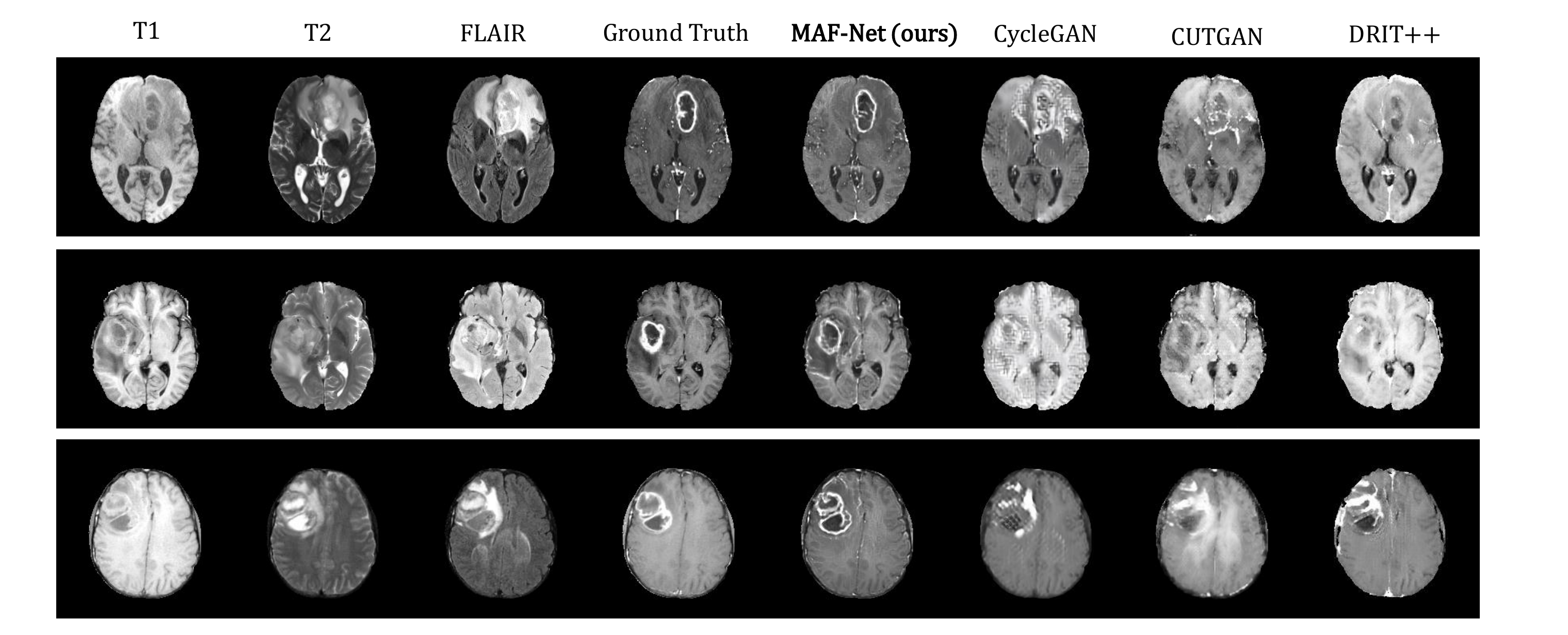}}
\vspace{-0.25cm}
\caption{Visualization of representative synthesized T1ce images.} 
\vspace{-0.25cm}
\label{fig2} 
\end{figure*}
\vspace{-0.0cm}

\vspace{+0.1cm}
\textbf{Ablation Study. }
An ablation study is conducted in Table \ref{ablation1} to evaluate our method's effectiveness. We perform simple averaging instead of MAF-Block to identify the effectiveness of our fusion component. The Dice score quantifies the overlap between the ground truth segmentation and the predicted segmentation and the average symmetric surface distance (ASSD) \cite{dice2015metrics} measures the average distance between the ground truth surface and the predicted surface. 

As shown in Table \ref{ablation1}, compared with the baseline UNet, which is fed with the three common modalities but not the synthesized T1ce, UNet together with a single encoder from CUTGAN improves both on the Dice scores and the ASSD metrics. Noticeably, with the incorporation of MAF-Block, significant increases are observed on the Dice scores of the enhancing tumor and the tumor core. The results reach the best when including all proposed components together with multi-task joint learning, revealing that joint learning produces better surfaces as verified by the ASSD values of the enhancing tumor and the tumor core. 


\vspace{+0.1cm}
\textbf{Modality Synthesis Results. }
We use Structure Similarity Index Measure (SSIM) and Peak Signal-to-Noise Ratio (PSNR) \cite{hore2010image} to measure the quality of the generated images and compare our proposed method with state-of-the-art image generation frameworks including CycleGAN \cite{Zhu2017UnpairedIT}, DRIT++ \cite{DRIT} and CUTGAN \cite{cutgan}. SSIM mainly considers three key features of an image: luminance, contrast and structure, while PSNR measures an image's quality. 

Fig. \ref{fig2} presents representative visualization results of all synthesized T1ce. Obviously, the T1ce images generated from MAF-Net, clearly distinguishing the enhancing tumor from the whole tumor, are more realistic and reasonable than those from other compared frameworks. 
It is worth mentioning that our method exhibits a more powerful localization ability of the tumor area, which is beneficial for the subsequent tumor segmentation task. 

Table \ref{tab1} quantitatively displays the comparison results of all methods, in terms of SSIM and PSNR. Apparently, our proposed MAF-Net largely outperforms other frameworks in synthesizing T1ce from a combination of T1, T2, and FLAIR. Compared with the second best framework CycleGAN, MAF-NET improves SSIM from 0.85 to 0.88 and PSNR from 21.62 to 22.78.

\begin{table}[h]
\centering
\caption{Quantitative comparisons of different methods on T1ce synthesis, in terms of SSIM and PSNR.}
\label{tab1}
\resizebox{5.5cm}{!}{
\begin{tabular}{lll}
\toprule
Method     & SSIM$\uparrow$  & PSNR$\uparrow$  \\ 
\midrule
DRIT++ \cite{DRIT}       &    0.8417           &  20.28             \\
CUTGAN \cite{cutgan}   &      0.8432     &        21.37       \\
CycleGAN \cite{Zhu2017UnpairedIT}     &      0.8574         &       21.62        \\
MAF-Net &       \textbf{0.8879}        &        \textbf{22.78}       \\ 
\bottomrule 
\end{tabular}}
\end{table}

\begin{figure}[htb]
\centering
\centerline{\includegraphics[width=9.5cm]{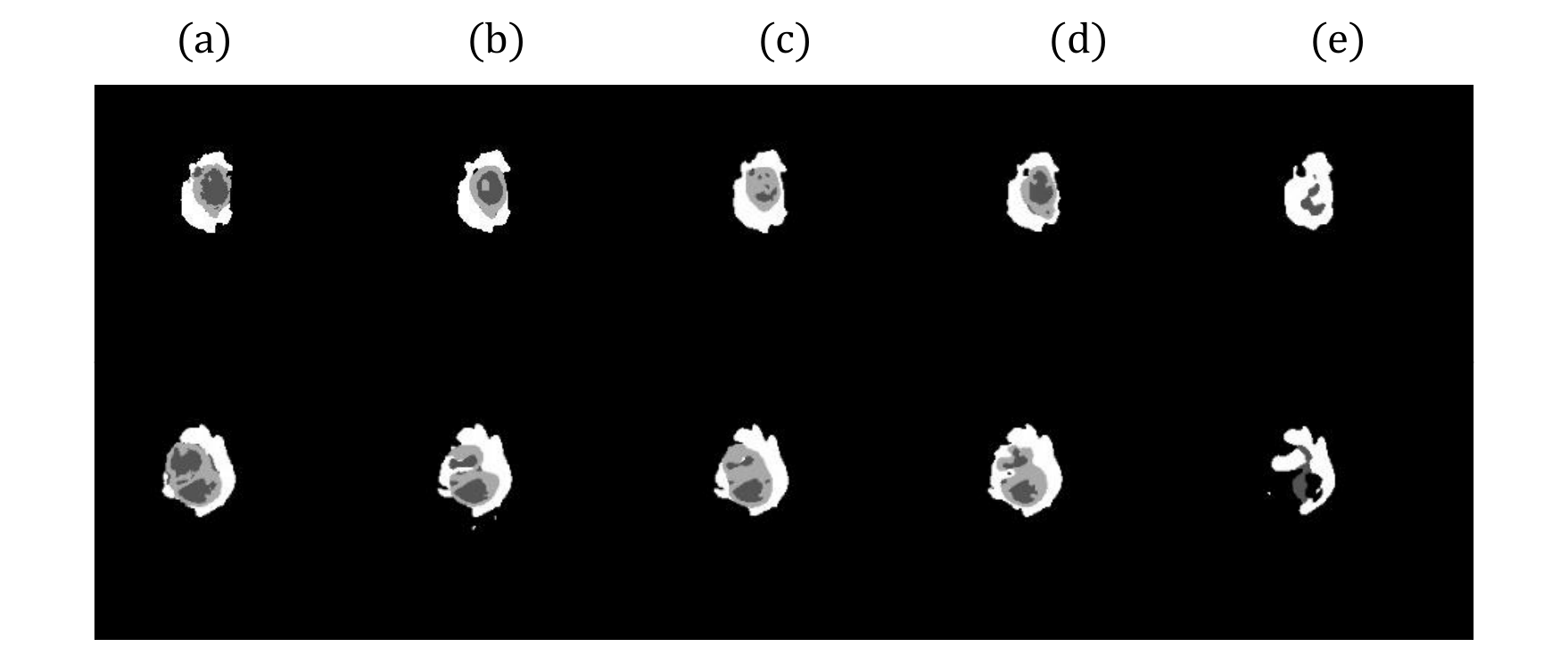}}
\vspace{-0.25cm}
\caption{Visualization of representative predicted mask and ground truth segmentations without post-processing. The white mask, gray mask and dark gray mask respectively denote edema, enhancing tumor and necrosis. (a) is the ground truth; (b)-(e) are the predicted segmentation masks from MAF-Net (ours), CycleGAN + UNet, CUTGAN + UNet and DRIT++ + UNet.} 
\vspace{-0.25cm}
\label{predictmask} 
\end{figure}

\textbf{Tumor Segmentation Results. }
 We add UNet to the aforementioned three image generation models (DRIT++, CUTGAN, and CycleGAN) so as to compare them with MAF-Net in terms of tumor segmentation in Table \ref{tab2} and quantitatively evaluate the results by the Dice score. For a fair comparison, we concatenate the three regular modalities and feed them into the image generation models. 
Fig. \ref{predictmask} presents representative segmentation results. Apparently, MAF-Net's results display more distinctive boundaries than those from other methods, especially for the necrosis and enhancing tumor. 
Referring to the quantitative evaluations in Table \ref{tab2}, it is clear that MAF-Net better benefits our downstream tumor segmentation task than each of the three compared frameworks, as clearly revealed by the Dice scores (67.9\% for the tumor core and 41.8\% for the enhancing tumor). 
Considering the results of modality synthesis, there is no doubt that high-quality synthesized T1ce affects the tumor segmentation performance in a good manner.
We also compare MAF-Net with other existing segmentation pipelines under the same missing modality setting. It is worth pointing out that most of the existing missing modality segmentation models require paired T1ce in the training phase, whereas our model does not. Results in Table \ref{tab2} illustrate that our MAF-Net compares favorably with models trained with paired samples, especially for segmenting the enhancing tumor and the tumor core.

\vspace{-0.35cm}
\begin{table}[htb]
\centering
\caption{The average Dice scores of different methods for brain tumor segmentation. Keys: WT -- whole tumor, ET -- enhancing tumor, TC -- tumor core. Brats2020 covers Brats19 and Brats18. }
\label{tab2}
\resizebox{8.5cm}{!}{
\begin{tabular}{llll}
\toprule
Method                      & \textbf{WT} & \textbf{ET} & \textbf{TC} \\ 
\midrule
\textbf{\emph{Paired T1ce for training}}    &               &                          &                     \\ 
Hamghalam. et al. (Brats19) \cite{hamghalam2021modality} & 88.0\%      & 34.1\%          & 61.5\%     \\
Zhou. et al. (Brats18) \cite{tip2021missing}      & 85.4\%      & 12.9\%          & 44.4\%  \\ 
\midrule
\textbf{\emph{Unpaired T1ce for training}}  &                      &                          &                     \\ 
DRIT++ + UNet                & 70.8\%               & 14.0\%                   & 33.5\%              \\
CycleGAN + UNet             & 87.5\%               & 38.5\%                   & 63.0\%              \\
CUTGAN + UNet               & 85.6\%               & 37.8\%                   & 62.5\%              \\ 
MAF-Net                  & \textbf{88.0\%}      & \textbf{41.8\%}          & \textbf{67.9\%}     \\ 
\bottomrule 
\end{tabular}}
\end{table}

\vspace{-0.15cm}
\section{Conclusion}
\label{sec:typestyle}
In this paper, we proposed and validated a novel multi-modal framework for unpaired missing modality synthesis and brain tumor segmentation. Utilizing patchwise contrastive loss and Modality-Level Attention Fusion, the proposed MAF-Net can effectively learn the interrelationship between each regular modality and the target T1ce, and dynamically attribute suitable weight to fuse multi-modal adaptive features. The decoder uses the fused features to generate the corresponding T1ce image and applies it to the downstream segmentation task. Our model can be trained with unpaired samples, which largely broadens the model's feasibility and applicability, and still achieves competitive performance in terms of both T1ce synthesis and brain tumor segmentation.


\section{Acknowledgement}
\label{sec:typestyle}
This study was supported by the National Natural Science Foundation of China (62071210); the Shenzhen Basic Research Program (JCYJ20200925153847004, JCYJ20190809120205578); the High-level University Fund (G02236002).
\bibliographystyle{IEEEtran}
\bibliography{GAN,cyclegan,cutgan,DRIT,autogan,TMI2020missing,drit_use_in_medical,cvpr2021translate,miccai2019dou,tip2021missing,ISBI2020,hamghalam2021modality,unet,SEmodule,brats,nce,ssim,dice,pix2pix,multimodal1,deepmultimodal1,deepmultimodal2,deepmultimodal3,deepmultimodal4,SVM,fusion1,fusion2,transfuse}

\begin{thebibliography}{10}
\providecommand{\url}[1]{#1}
\csname url@samestyle\endcsname
\providecommand{\newblock}{\relax}
\providecommand{\bibinfo}[2]{#2}
\providecommand{\BIBentrySTDinterwordspacing}{\spaceskip=0pt\relax}
\providecommand{\BIBentryALTinterwordstretchfactor}{4}
\providecommand{\BIBentryALTinterwordspacing}{\spaceskip=\fontdimen2\font plus
\BIBentryALTinterwordstretchfactor\fontdimen3\font minus
  \fontdimen4\font\relax}
\providecommand{\BIBforeignlanguage}[2]{{%
\expandafter\ifx\csname l@#1\endcsname\relax
\typeout{** WARNING: IEEEtran.bst: No hyphenation pattern has been}%
\typeout{** loaded for the language `#1'. Using the pattern for}%
\typeout{** the default language instead.}%
\else
\language=\csname l@#1\endcsname
\fi
#2}}
\providecommand{\BIBdecl}{\relax}
\BIBdecl

\bibitem{miccai2019dou}
C.~Chen, Q.~Dou, Y.~Jin \emph{et~al.}, ``Robust multimodal brain tumor
  segmentation via feature disentanglement and gated fusion,'' in
  \emph{International Conference on Medical Image Computing and
  Computer-Assisted Intervention}.\hskip 1em plus 0.5em minus 0.4em\relax
  Springer, 2019, pp. 447--456.

\bibitem{tip2021missing}
T.~Zhou, S.~Canu, P.~Vera \emph{et~al.}, ``Latent correlation representation
  learning for brain tumor segmentation with missing mri modalities,''
  \emph{IEEE Transactions on Image Processing}, vol.~30, pp. 4263--4274, 2021.

\bibitem{hamghalam2021modality}
M.~Hamghalam, A.~F. Frangi, B.~Lei \emph{et~al.}, ``Modality completion via
  gaussian process prior variational autoencoders for multi-modal glioma
  segmentation,'' \emph{arXiv preprint arXiv:2107.03442}, 2021.

\bibitem{cvpr2021translate}
Z.~Zhang, L.~Yang, and Y.~Zheng, ``Translating and segmenting multimodal
  medical volumes with cycle-and shape-consistency generative adversarial
  network,'' in \emph{Proceedings of the IEEE conference on computer vision and
  pattern Recognition}, 2018, pp. 9242--9251.

\bibitem{multimodal1}
B.~H. Menze, K.~Van~Leemput, D.~Lashkari \emph{et~al.}, ``A generative model
  for brain tumor segmentation in multi-modal images,'' in \emph{International
  Conference on Medical Image Computing and Computer-Assisted
  Intervention}.\hskip 1em plus 0.5em minus 0.4em\relax Springer, 2010, pp.
  151--159.

\bibitem{deepmultimodal1}
J.~Dolz, K.~Gopinath, J.~Yuan,  \emph{et~al.}, ``Hyperdense-net: {A}
  hyper-densely connected {CNN} for multi-modal image segmentation,''
  \emph{CoRR}, vol. abs/1804.02967, 2018.

\bibitem{deepmultimodal2}
X.~Feng, N.~J. Tustison, S.~H. Patel \emph{et~al.}, ``Brain tumor segmentation
  using an ensemble of 3d u-nets and overall survival prediction using radiomic
  features,'' \emph{Frontiers in Computational Neuroscience}, vol.~14, 2020.

\bibitem{deepmultimodal4}
R.~Hua, Q.~Huo, Y.~Gao \emph{et~al.}, ``Segmenting brain tumor using cascaded
  v-nets in multimodal mr images,'' \emph{Frontiers in computational
  neuroscience}, vol.~14, p.~9, 2020.

\bibitem{deepmultimodal3}
R.~Ranjbarzadeh, A.~Bagherian~Kasgari, S.~Jafarzadeh~Ghoushchi \emph{et~al.},
  ``Brain tumor segmentation based on deep learning and an attention mechanism
  using mri multi-modalities brain images,'' \emph{Scientific Reports},
  vol.~11, no.~1, p. 10930, May 2021.

\bibitem{goodfellow2014generative}
I.~Goodfellow, J.~Pouget-Abadie, M.~Mirza \emph{et~al.}, ``Generative
  adversarial nets,'' \emph{Advances in neural information processing systems},
  vol.~27, 2014.

\bibitem{pix2pix}
P.~Isola, J.-Y. Zhu, T.~Zhou, and A.~A. Efros, ``Image-to-image translation
  with conditional adversarial networks,'' in \emph{Proceedings of the IEEE
  conference on computer vision and pattern recognition}, 2017, pp. 1125--1134.

\bibitem{Zhu2017UnpairedIT}
J.-Y. Zhu, T.~Park, P.~Isola \emph{et~al.}, ``Unpaired image-to-image
  translation using cycle-consistent adversarial networks,'' \emph{2017 IEEE
  International Conference on Computer Vision (ICCV)}, pp. 2242--2251, 2017.

\bibitem{cutgan}
T.~Park, A.~A. Efros, R.~Zhang \emph{et~al.}, ``Contrastive learning for
  unpaired image-to-image translation,'' in \emph{European Conference on
  Computer Vision}.\hskip 1em plus 0.5em minus 0.4em\relax Springer, 2020, pp.
  319--345.

\bibitem{fusion2}
T.~Zhou, K.-H. Thung, M.~Liu, F.~Shi, C.~Zhang, and D.~Shen, ``Multi-modal
  latent space inducing ensemble svm classifier for early dementia diagnosis
  with neuroimaging data,'' \emph{Medical Image Analysis}, vol.~60, p. 101630,
  2020.

\bibitem{SVM}
C.~Cortes and V.~Vapnik, ``Support-vector networks,'' \emph{Machine learning},
  vol.~20, no.~3, pp. 273--297, 1995.

\bibitem{fusion1}
L.~Chen, Y.~Wu, A.~M. DSouza, A.~Z. Abidin, A.~Wismuller, and C.~Xu, ``Mri
  tumor segmentation with densely connected 3d cnn,'' 2018.

\bibitem{SEmodule}
J.~Hu, L.~Shen, and G.~Sun, ``Squeeze-and-excitation networks,'' in
  \emph{Proceedings of the IEEE conference on computer vision and pattern
  recognition}, 2018, pp. 7132--7141.

\bibitem{transfuse}
A.~Prakash, K.~Chitta, and A.~Geiger, ``Multi-modal fusion transformer for
  end-to-end autonomous driving,'' in \emph{Proceedings of the IEEE/CVF
  Conference on Computer Vision and Pattern Recognition}, 2021, pp. 7077--7087.

\bibitem{drit_use_in_medical}
J.~Yang, N.~C. Dvornek, F.~Zhang \emph{et~al.}, ``Unsupervised domain
  adaptation via disentangled representations: Application to cross-modality
  liver segmentation,'' in \emph{International Conference on Medical Image
  Computing and Computer-Assisted Intervention}.\hskip 1em plus 0.5em minus
  0.4em\relax Springer, 2019, pp. 255--263.

\bibitem{TMI2020}
A.~Sharma and G.~Hamarneh, ``Missing mri pulse sequence synthesis using
  multi-modal generative adversarial network,'' \emph{IEEE Transactions on
  Medical Imaging}, vol.~39, no.~4, pp. 1170--1183, 2020.

\bibitem{isbi2020}
B.~Xin, Y.~Hu, Y.~Zheng \emph{et~al.}, ``Multi-modality generative adversarial
  networks with tumor consistency loss for brain mr image synthesis,'' in
  \emph{2020 IEEE 17th International Symposium on Biomedical Imaging
  (ISBI)}.\hskip 1em plus 0.5em minus 0.4em\relax IEEE, 2020, pp. 1803--1807.

\bibitem{oord2018representation}
A.~v.~d. Oord, Y.~Li, and O.~Vinyals, ``Representation learning with
  contrastive predictive coding,'' \emph{arXiv preprint arXiv:1807.03748},
  2018.

\bibitem{unet}
O.~Ronneberger, P.~Fischer, and T.~Brox, ``U-net: Convolutional networks for
  biomedical image segmentation,'' in \emph{International Conference on Medical
  image computing and computer-assisted intervention}.\hskip 1em plus 0.5em
  minus 0.4em\relax Springer, 2015, pp. 234--241.

\bibitem{brats}
B.~H. Menze, A.~Jakab, S.~Bauer \emph{et~al.}, ``The multimodal brain tumor
  image segmentation benchmark (brats),'' \emph{IEEE transactions on medical
  imaging}, vol.~34, no.~10, pp. 1993--2024, 2014.

\bibitem{dice2015metrics}
A.~A. Taha and A.~Hanbury, ``Metrics for evaluating 3d medical image
  segmentation: analysis, selection, and tool,'' \emph{BMC medical imaging},
  vol.~15, no.~1, pp. 1--28, 2015.

\bibitem{hore2010image}
A.~Hore and D.~Ziou, ``Image quality metrics: Psnr vs. ssim,'' in \emph{2010
  20th international conference on pattern recognition}.\hskip 1em plus 0.5em
  minus 0.4em\relax IEEE, 2010, pp. 2366--2369.

\bibitem{DRIT}
H.-Y. Lee, H.-Y. Tseng, Q.~Mao \emph{et~al.}, ``Drit++: Diverse image-to-image
  translation via disentangled representations,'' \emph{International Journal
  of Computer Vision}, vol. 128, no.~10, pp. 2402--2417, 2020.

\end{thebibliography}

%



\end{document}